\documentclass[aps,prl,twocolumn,showpacs,superscriptaddress,groupedaddress,nofootinbib]{revtex4}
\usepackage[utf8x]{inputenc}
\usepackage{graphicx}
\usepackage{amsmath,amssymb} 
\usepackage[english]{babel}
\usepackage{rotating}
\usepackage{amsbsy}
\usepackage{textcomp}
\usepackage{psfrag}
\usepackage{multirow}
\usepackage[usenames]{color} 
\usepackage{sidecap}
\usepackage{array}
\usepackage{xspace}
\usepackage{subfig}
\usepackage{natbib}
\usepackage{microtype}
\usepackage{xcolor,ulem}
\usepackage{hyperref}
\usepackage{acronym}
\usepackage{layouts}

\newacro{GW}{gravitational wave}
\newacro{IMBHB}{intermediate-mass black hole binary}

\newcommand{\GW}{\ac{GW}}
\newcommand{\IMBHB}{\ac{IMBHB}}

\newcommand{\referee}[1]{{#1}}

\newcommand{\Msun}{\ensuremath{\mathrm{M}_\odot}}
\newcommand{\Mc}{\ensuremath{\mathcal{M}}}
\newcommand{\Mtot}{\ensuremath{M_\mathrm{total}}}

\def\apjl{Astrophysical Journal}             

\def\apjs{ApJS}              
\def\aap{A\&A}                

\pacs{04.30.Tv,97.60.Lf}

\begin{document}
\title{Measuring intermediate mass black hole binaries with advanced gravitational wave detectors}

\author{John Veitch}
\affiliation{School of Physics and Astronomy, University of Birmingham, Birmingham, B15 2TT, United Kingdom}

\author{Michael P\"{u}rrer}
\affiliation{School of Physics and Astronomy, Cardiff University, Cardiff, CF24 3AA, United Kingdom}

\author{Ilya Mandel}
\affiliation{School of Physics and Astronomy, University of Birmingham, Birmingham, B15 2TT, United Kingdom}
\affiliation{Monash Center for Astrophysics, Monash University, Clayton, VIC 3800, Australia}

\begin{abstract}
        We perform a systematic study to explore the accuracy with which the parameters of  \IMBHB\ systems can be measured from their \GW\ signatures using second-generation \GW\ detectors. We make use of the most recent reduced-order models containing inspiral, merger and ringdown signals of aligned-spin effective-one-body waveforms (SEOBNR) to significantly speed up the calculations.  We explore the phenomenology of the measurement accuracies for binaries with total masses between $50$ and $500$\,\Msun\ and mass ratios between $0.1$ and $1$.  We find that (i) at total masses below $\sim 200$\,\Msun, where the signal-to-noise-ratio is dominated by the inspiral portion of the signal, the chirp mass parameter can be accurately measured; (ii) at higher masses, the information content is dominated by the ringdown, and total mass is measured more accurately; (iii) the mass of the lower-mass companion is poorly estimated, especially at high total mass and more extreme mass ratios; (iv) spin cannot be accurately measured for our injection set with non-spinning components.  Most importantly, we find that for binaries with non-spinning components at all values of the mass ratio in the considered range and at network signal-to-noise ratio of $15$, analyzed with spin-aligned templates, the presence of an intermediate-mass black hole with mass $> 100$\,\Msun\ can be confirmed with 95\% confidence in any binary that includes a component with a mass of $130$\,\Msun\ or greater.
\end{abstract}

\maketitle

\section{Introduction}\label{s:intro}

Advanced LIGO \cite{AdvLIGO} and Virgo \cite{AdvVirgo} detectors are expected to start taking data in late 2015 and 2016 \cite{scenarios}, respectively.  Compact binary coalescences are a key source of gravitational-wave (GW) signals for advanced detectors \cite[e.g.,][]{ratesdoc,MandelOShaughnessy:2010}.  These may include binaries where one or both components are intermediate-mass black holes (IMBHs), with mass in the $\sim 50$ --- few hundred \Msun\ range. 

There is growing observational and theoretical evidence for the existence of IMBHs in globular clusters \cite[see][for review]{MillerColbert:2004}.  Observational evidence comes in the form of observations of ultra-luminous X-ray sources (ULX), cluster dynamics (though these are mostly sensitive to higher-mass IMBHs, whose GW signatures would be at frequencies below the detectors' sensitive band), and, most recently, a tentative quasi-periodic oscillator observation of a 400 \Msun\ IMBH \cite{Pasham:2014}.  On the theoretical side, a number of models have been predicted for IMBH growth, from direct collapse from very massive stars \cite[e.g.,][]{MadauRees:2001} to runaway collision scenarios \cite{PZwart:2002,PZwart:2004} or gradual growth through stellar-mass BH mergers \cite[e.g.,][]{OLeary:2006} or accretion \cite{Vesperini:2010}. 

Advanced GW detectors could observe inspirals of stellar-mass compact-objects into IMBHs in globular clusters \cite{Mandel:2008}.  IMBH binary mergers are possible if the binary fraction in a globular cluster is sufficiently high to allow the formation of two IMBHs \cite{Fregeau:2006}, or via mergers of two globular clusters with each other and subsequent coalescences of the IMBHs they host \cite{Amaro:2006imbh,AmaroSeoaneSantamaria:2009}.  Outside of globular clusters, merging compact binaries including IMBHs could form directly from isolated binaries composed of very massive ($\gtrsim 300$\,\Msun) stars \cite{Belczynski:2014VMS}.  All of these scenarios could produce advanced-detector event rates of tens or more detections per year, though much lower rates are also possible.  Meanwhile, future detectors with good low-frequency sensitivity, such as the proposed Einstein Telescope \cite{ET}, could observe up to thousands of IMBH binary mergers per year \cite{Gair:2009ETrev} and use high-redshift IMBH binary observations to probe the history of massive black hole formation \cite{Sesana:2009ET,Gair:2009ET}.

GW observations, which allow for a direct mass measurement, could provide the first completely unambiguous proof of the existence of IMBHs in the few-hundred-solar-mass range.  If such IMBHs are discovered, their observations would shed light on very massive star evolution and globular cluster dynamics.  IMBHs could also prove to be particularly accurate probes of strong-field dynamical gravity, allowing for tests of the general theory of relativity \cite[e.g.,][]{Brown:2007,Rodriguez:2012}. \referee{As the coalescence of IMBHBs is expected to be electromagnetically quiet, gravitational waves are likely to be the only means of observing these systems directly.} For this reason, the LIGO and Virgo collaborations have carried out searches for IMBH binaries in initial detector data (which did not yield detections) \cite{IMBHB:S5,IMBHB:S6,Ringdown:S5S6} and intend to do the same in the advanced detector era with more sensitive instruments.  However, in order to establish that an IMBH has been detected and to explore the scientific consequences of this discovery, it is necessary to analyze the GW signature of a coalescence in order to infer the parameters of the systems, particularly the component masses. \referee{In this paper we perform the first systematic study of the accuracy of \IMBHB\ mass measurements achievable with GW observations.}

Accurate models for \acp{GW} emitted from IMBH binaries must include the inspiral, merger, and ringdown phases of the coalescence.  One of the most accurate available theoretical waveform families are effective-one-body (EOB) models~\cite{Buonanno:1998gg,Buonanno:2000ef,Damour:2001tu}. EOB is an analytical approach that combines post-Newtonian expansion, re-summation techniques and perturbation theory with additional calibration of certain model parameters against waveforms obtained by numerically integrating Einstein's equations on super-computers. These models are defined via a complicated set of ordinary differential equations in the time-domain and can be very computationally expensive to generate, limiting their use in parameter estimation studies to date. Novel reduced order modeling (ROM) techniques~\cite{Field:2013cfa,Puerrer:2014fza,Purrer:2015tud} have allowed for the construction of fast and accurate surrogate models of EOB waveforms. In particular, we use the frequency domain ROMs~\cite{Puerrer:2014fza,Purrer:2015tud} for \referee{EOB waveforms with spins aligned with the orbital angular momentum of the binary}, defined in~\cite{Taracchini:2012ig,Taracchini:2013rva} and implemented in LAL~\cite{LAL-web}. \referee{This allows us to perform simulations on a scale unprecedented for this class of sources}.

\section{Simulations}\label{s:simulations}
\begin{figure}
    \includegraphics{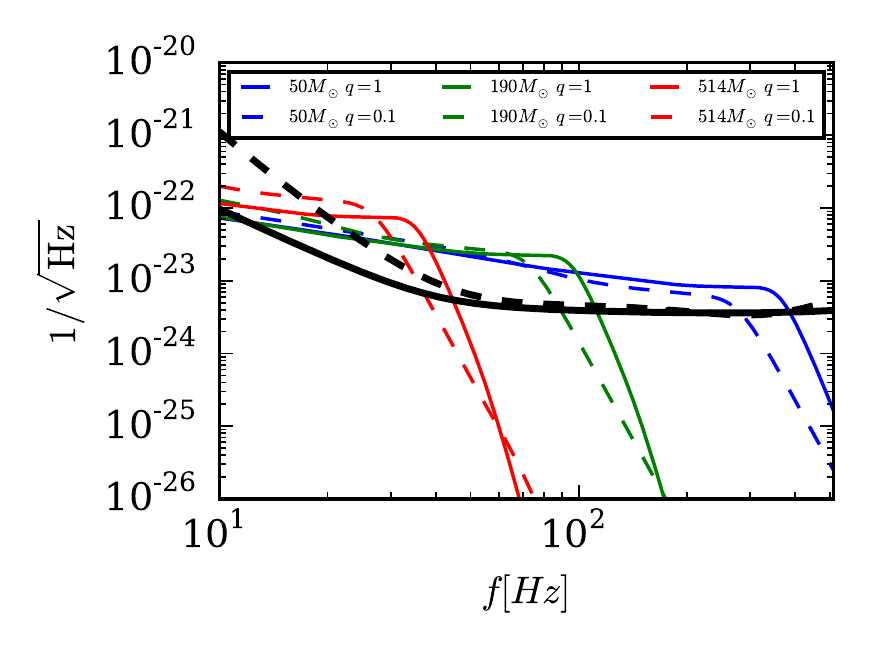}
    \caption{\label{fig:Ringdown}
Characteristic amplitudes $h_c \equiv \sqrt{f} \lvert \tilde h(f) \rvert$ of SEOBNRv2 \cite{Taracchini:2013rva} injections with various total masses and mass ratios, at network SNR 15 used in this study. In black, the detector noise amplitude spectrum $\sqrt{S_n(f)}$ of the Advanced LIGO design noise-curve (solid) and the Advanced VIRGO noise-curve (dashed).
}
\end{figure}

We performed a systematic study of the accuracy with which the masses and spins of the IMBHB could be recovered from \GW\ observations by Advanced LIGO and Advanced Virgo \GW\ detectors operating at design sensitivity. For Advanced LIGO we used the zero-detuned, high-power detector configuration~\cite{LIGO-T1200307}, and for Advanced Virgo a phenomenological fit to the design sensitivity curve~\cite{ManzottiDietz:2012}, both of which are displayed in Figure~\ref{fig:Ringdown} along with typical characteristic signal amplitudes. For our analysis we began generating the waveforms at a lower frequency of 10\,Hz in both LIGO and Virgo instruments, allowing us to take full advantage of the low-frequency sensitivity of the instruments, which will be achieved toward the end of the decade.  The use of zero-noise mock data sets to estimate parameter measurement accuracy relies on the assumption that the noise is stationary and Gaussian; although Berry et al.~\cite{Berry:2015} demonstrated that non-stationary realistic noise does not significantly influence parameter estimation for neutron star binaries, departures from stationarity (noise ``glitches'') could play a larger role for low-frequency, short-duration IMBHB signals.

To investigate parameter measurement accuracy, we analyzed a set of mock data sets (injections) with the \texttt{LALInference} \cite{LALInference} Bayesian parameter-estimation pipeline.  This pipeline returns a set of samples from the joint posterior distribution for the signal parameters.  We can readily convert this output into the innermost 90\% credible region, spanning from the 5th to the 95th percentile, on marginalized single-parameter posterior distributions, which we use as a proxy for measurement accuracy.  We find this to be a more robust metric for measurement accuracy than the standard deviation of the highly non-Gaussian posteriors.

We injected data with both SEOBNRv1 and SEOBNRv2 \cite{Taracchini:2013rva} waveforms, and used corresponding single-spin ROM template families for recovery.  We found resulting measurement accuracies that are qualitatively and quantitatively similar, so we only show results from the more recent SEOBNRv2 model here.  We included only the dominant $l=m=2$ mode of the gravitational wave signal in our simulations, as this is the only mode included in the SEOBNRv2 \referee{reduced order} model. This omission means that our results can only serve as a conservative estimate of the parameter estimation performance for \IMBHB\ systems, since the higher frequency harmonics of the signal can carry information to further constrain the signal model, especially at high masses~\cite{Littenberg:2012uj,Varma:2014jxa,Graff:2015}. \referee{Further development of ROMs to include both higher harmonics and spin is necessary to provide timely results with the best possible accuracy}. We did not include the cosmological redshift of the waveforms, so our results should be interpreted as measurements of the redshifted masses in the rest frame of the detectors.

In order to explore measurement accuracy as a function of mass and mass ratio, we carried out injections for a broad range of total masses $\Mtot=m_1+m_2$ between 50 and 500\,\Msun, where $m_1>m_2$ are the component masses.  For each mass, we injected systems at four mass ratios $q=m_2/m_1$ of 1, 1/2, 1/4 and 1/10, always with non-spinning components.  All simulated signals were oriented such that the orbital angular momentum vector was inclined at $30^\circ$ to the line of sight between the Earth and the binary.  Although the orientation and sky location were the same in all simulations, we do not expect the measurement of the mass and aligned spin parameters to be significantly affected by this choice\referee{, since we do not include higher modes which can couple mass ratio measurement to extrinsic parameter accuracy~\cite{Graff:2015}}.  The distance was chosen to yield a constant coherent signal-to-noise ratio (SNR) of $15$.  

For the Bayesian analysis, we used flat priors on the component masses within the range $m_1, m_2 \in [5, 1000]$\,\Msun, limited the total mass to $\Mtot \le 1000$\,\Msun, and limited the mass ratio $m_2/m_1\geq 0.01$.   We did not assume any of the source parameters were known when performing parameter estimation, allowing an isotropic prior on orientation, and a uniform-in-volume prior on binary location out to a luminosity distance of $15$\,Gpc \referee{(a redshift of $\sim 1.9$)}.  The prior on the single aligned spin $\chi$ was fixed to be flat in $[-1, 0.6]$, the range of validity of the SEOBNRv1 \cite{Taracchini:2012ig} approximant.
        \referee{Since this prior distribution does not match the distribution of sources analyzed, we should anticipate that posteriors on individual injections can be centered away from the true values, despite the self-consistency of \texttt{LALInference}, which has been demonstrated to produce $X\%$ credible intervals that contain the true value $X\%$ of the time \cite{Sidery:2013, Berry:2015, LALInference}.  For example, the low a {\it a priori} probability of high-mass extreme-mass ratio injections with non-spinning components, coupled to the asymmetry in the impact of remnant spin on the well-measured central frequency of the dominant ringdown harmonic \cite[e.g.,][]{Berti:2006}, will lead to a typical over-estimate of the inferred total mass for such sources. This is compounded by the prior on distance $p(d_L)\propto d_L^2$, which for a fixed amplitude tends to prefer higher mass sources at greater distances.}

\section{Results}\label{s:results}

\paragraph{Mass measurement}

\begin{figure}
        \includegraphics{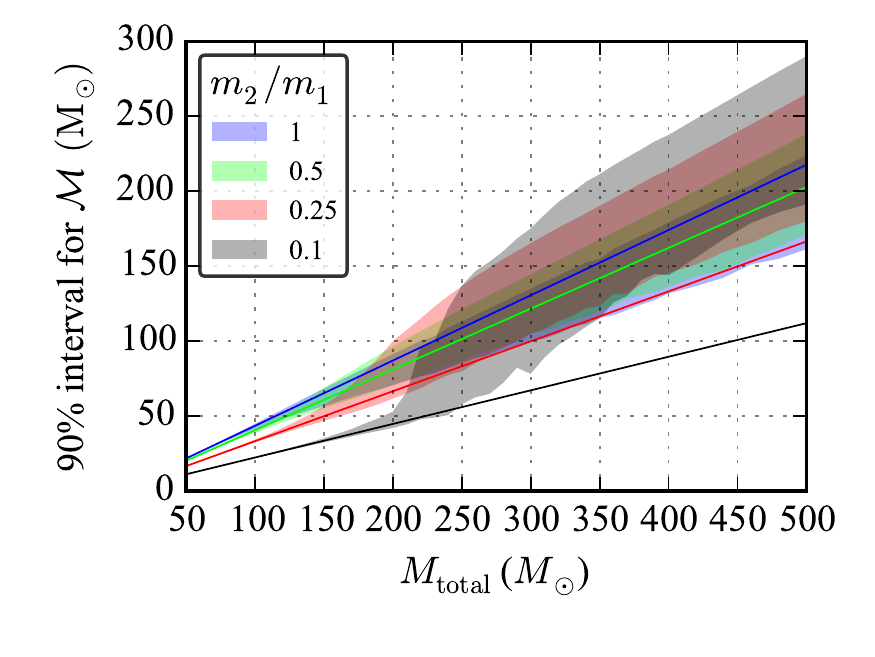}
        \caption{\label{fig:Mchirp}The 90\% credible intervals for the chirp mass \Mc\ as a function of total mass \Mtot, for four mass ratios $m_2/m_1$. True values are indicated by the solid lines.
        As \Mtot\ increases in the range $50-200$\,\Msun, the measurement of \Mc\ becomes steadily worse as the inspiral potion of the signal moves out of the sensitive band of the detector.}
\end{figure}

Figure \ref{fig:Mchirp} shows the $90\%$ credible interval for the chirp mass, $\Mc=m_1^{3/5} m_2^{3/5} \Mtot^{-1/5}$, as a function of the total mass \Mtot.

At lower masses, the signal is dominated by the `chirping' inspiral portion, and the phase evolution is a function of \Mc\ at leading order, which is therefore the most strongly constrained parameter when analysing lower mass systems~\cite{S6PE}. \referee{We find that the 
width of the 90\% credible interval on $\Mc$ is $0.3 - 0.5 M_\odot$ at $\Mtot = 50\Msun$ and $0.7-3.5 M_\odot$ at $M_{\rm tot}=100\Msun$.  For comparison, the same interval is typically $\lesssim 0.01\, /\, 0.03\, /\, 0.1 M_\odot$ for binary neutron star systems / neutron star -- black hole binaries / stellar-mass binary black holes, respectively~\cite[e.g.,][]{RodriguezEtAl:2013,S6PE,Mandel:2015}.}

\begin{figure}
    \includegraphics{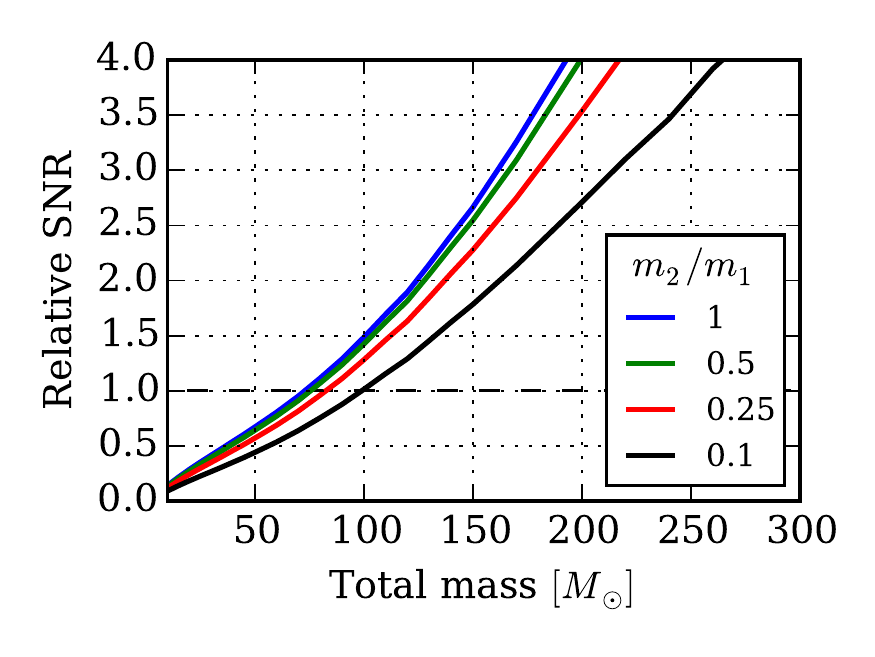}
    \caption{\label{fig:RelativeSNR}
The relative SNR, the ratio of the SNR above and the SNR below the GW frequency at the innermost stable circular orbit (ISCO). We use the Schwarzschild ISCO $f_\mathrm{ISCO} = 6^{-3/2} / (\pi M)$ which is strictly speaking only valid in the test particle limit. The relative SNR w.r.t.~the ISCO depends on the mass-ratio. In contrast, the ratio of SNRs above and below half the ringdown frequency of each system is only weakly dependent on the mass ratio and reaches unity at a total mass of $150$\,\Msun.
}
\end{figure}

\begin{figure}
        \includegraphics{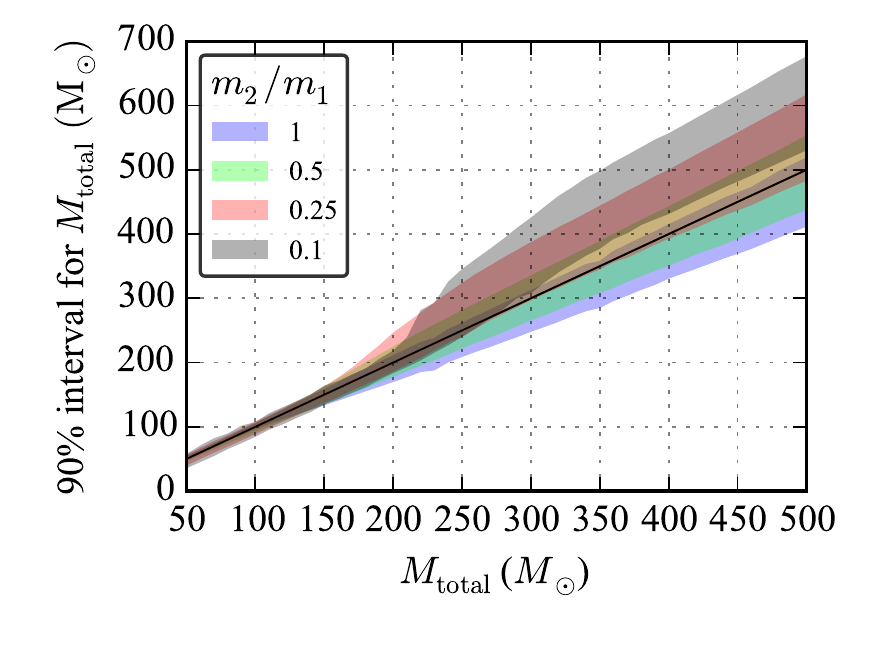}
        \caption{\label{fig:Mtot}The  90\% credible intervals for \Mtot.
}
\end{figure}

Meanwhile, as the mass increases, the inspiral moves to progressively lower frequencies and out of the sensitive band of the detector (see Figure \ref{fig:Ringdown}) and the merger and ringdown contribute an increasing fraction of the SNR (see Figure \ref{fig:RelativeSNR}).  At masses above $\sim 100$\,\Msun, the SNR is dominated by the merger and ringdown, and above $\sim 200$\,\Msun, by the ringdown.  The ringdown frequency depends only on the total mass and spin of the merger product (the latter is a function of the mass ratio for non-spinning components).  We therefore expect the total mass of high-mass systems to be better constrained than the chirp mass \cite[this has previously been pointed out by Graff, Buonanno, and Sathyaprakash in ref.][]{2015GReGr..47...11A}; moreover, the accuracy of the \Mtot\ measurement should become increasingly insensitive to the mass ratio.  Indeed, this is the behavior we see in Fig.~\ref{fig:Mtot}, which shows the $90\%$ credible interval for the total mass.

\begin{figure*}
        \includegraphics{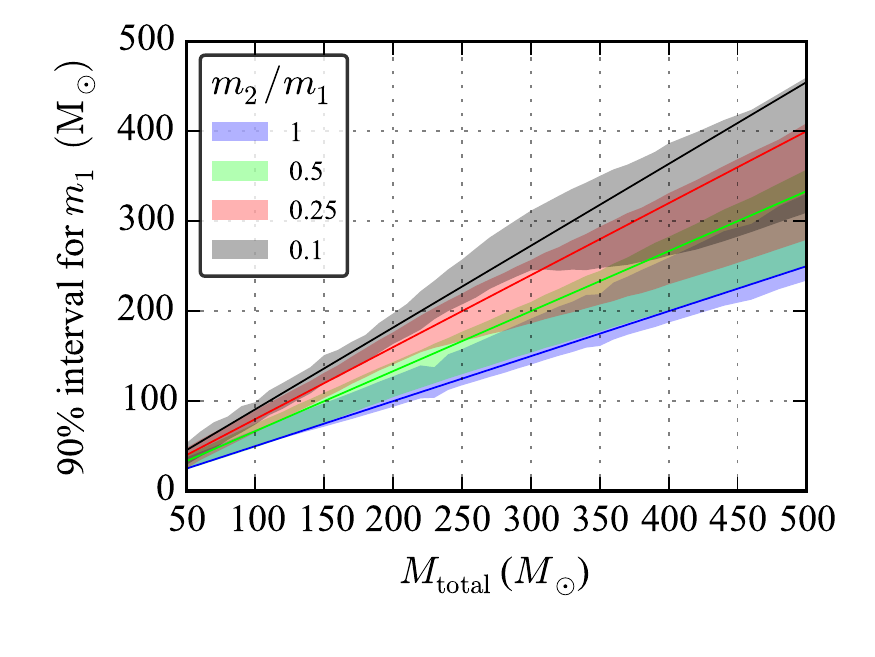}
        \includegraphics{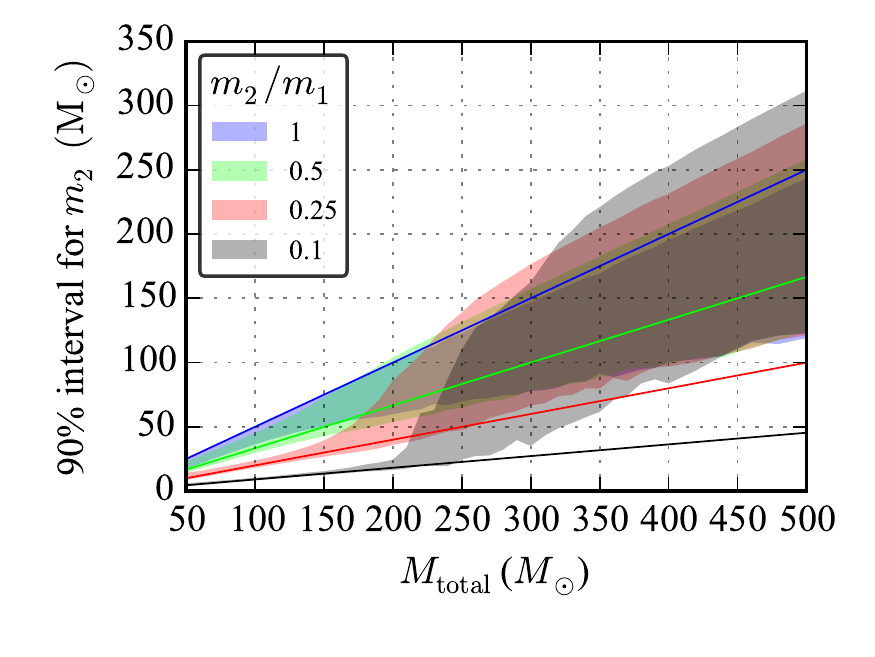}
        \caption{\label{fig:components}The 90\% credible intervals for the component masses $m_1$ (left, larger companion) and $m_2$ (right, smaller companion).
}
\end{figure*}

Alternatively, the mass measurement accuracy can be visualized by considering the 90\% credible region widths on component masses.  Figure \ref{fig:components} shows that component masses are generally harder to measure because of the significant uncertainty in the mass ratio typical for GW parameter estimation \cite{S6PE}.  The mass of the lower-mass component $m_2$ is particularly poorly constrained, especially at high masses and more extreme mass ratios, where only the total mass is encoded in the ringdown signature. \referee{The fractional uncertainty of the better-measured $m_1$ component varies between $40\%$ and $10\%$ between \Mtot\ 100 and 300\Msun.}

\begin{figure}
    \includegraphics{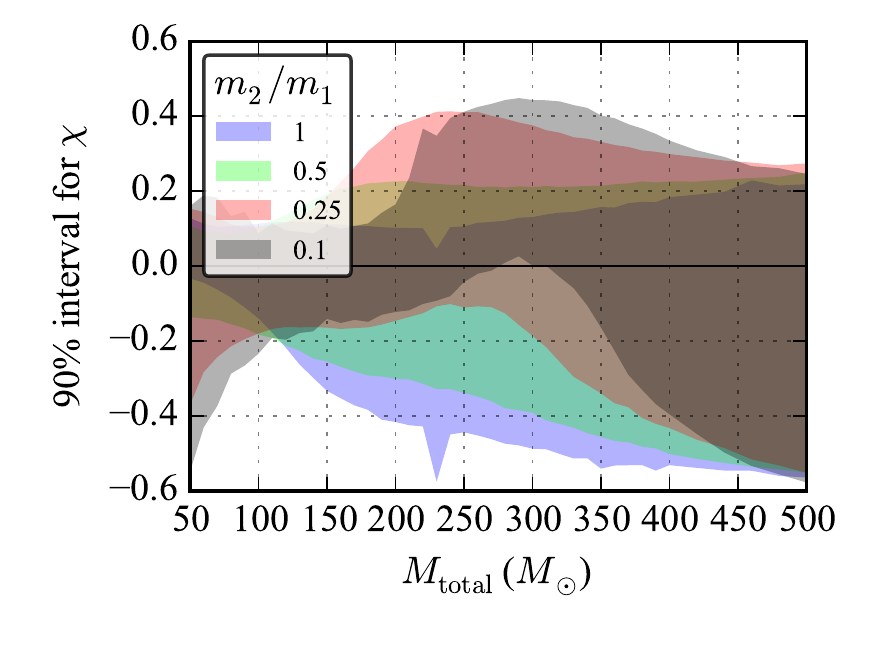}
    \caption{\label{fig:chi}The 90\% credible interval for the effective dimensionless spin $\chi$.}
\end{figure}

\paragraph{Spin}
All of our injections have non-spinning components, and the \referee{SEOBNRv2ROM waveform model which we used} includes only a single spin parameter $\chi = (m_1 \chi_1 + m_2 \chi_2) / M$, a combination of the dimensionless spins $\chi_i = \vec L \cdot \vec S_i / m_i^2$ aligned with the orbital angular momentum $\vec L$ that plays a dominant role in governing the inspiral phase evolution through spin-orbit coupling~\cite{Ajith:2011ec,Purrer:2013ojf}.
Figure \ref{fig:chi} shows that the measurement accuracy of $\chi$ decreases with total mass, as the inspiral moves out of the detector band.  In general, $\chi$ is not well constrained for non-spinning injections ($\chi=0$), as $\chi$ values between $\sim 0.2$ and $\sim -0.5$ are allowed, spanning about half of the prior range $[-1, 0.6]$.

\begin{figure}
    \includegraphics{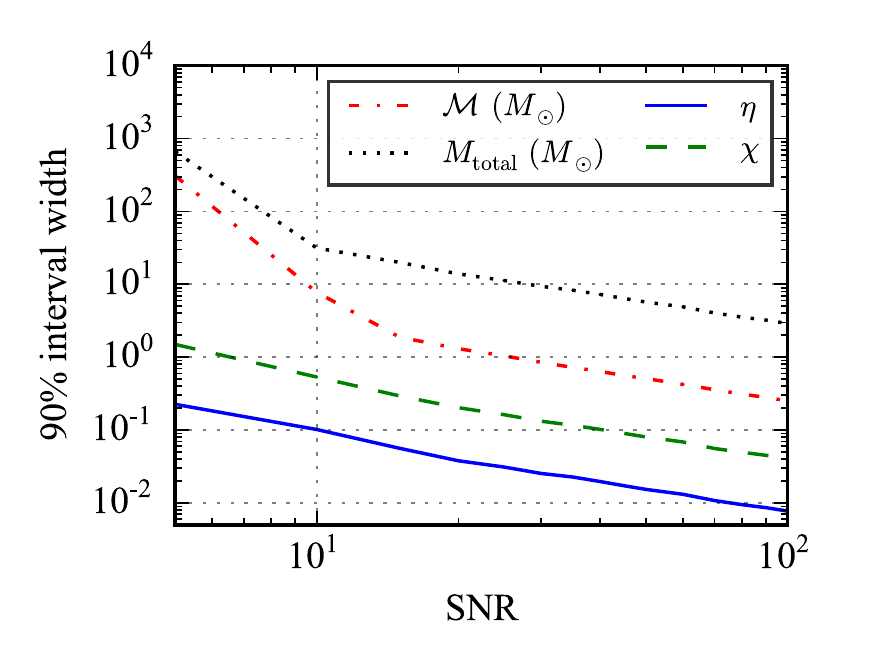}
    \caption{\label{fig:SNR}The width of the $90\%$ credible intervals in chirp mass \Mc, total mass \Mtot, symmetric mass-ratio $\eta$ and spin $\chi$ as a function of network SNR for mass-ratio $q=0.25$ and total mass $100$\,\Msun.
}
\end{figure}

\paragraph{Measurability of parameters as a function of signal-to-noise ratio}
We also performed a series of simulations where we increased the signal-to-noise ratio from 5 to 100. The shape of the posterior probability density function approaches a multivariate Gaussian at high SNR; once this happens, we expect uncertainties on individual parameters to fall off as SNR$^{-1}$.  As shown in Figure \ref{fig:SNR} this is indeed the case for SNRs larger than $\sim 15$.

\section{Discussion}\label{s:summary}
\begin{figure}
    \includegraphics{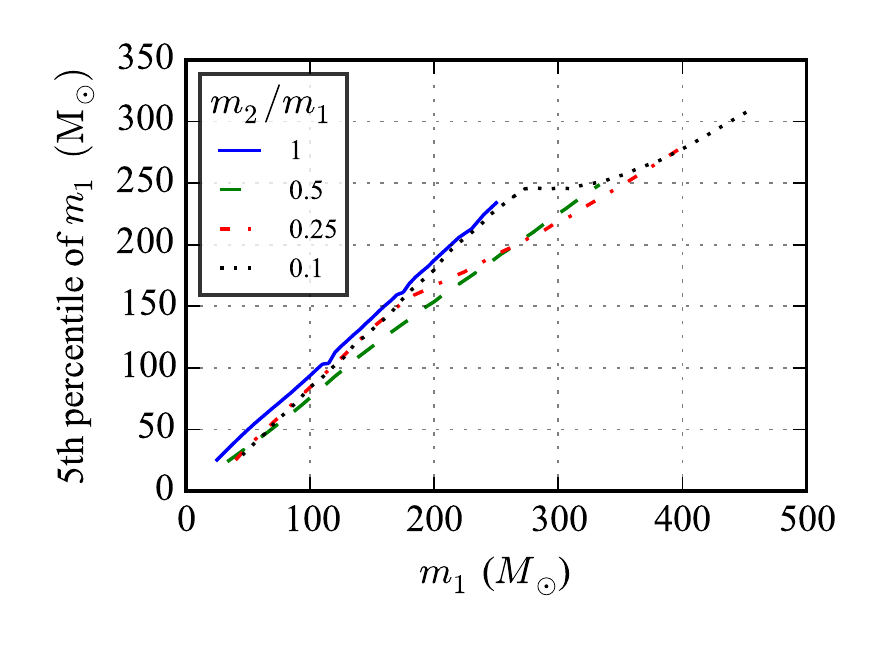}
    \caption{\label{fig:m1lowerbound}The 95\%--confidence lower bound on the mass of the more massive component $m_1$ as a function of $m_1$, showing that a system can be confidently classified as containing an IMBH with mass $>100$\,\Msun\ when $m_1$ exceeds $\sim130$\,\Msun.}
\end{figure}

A key question that will arise when a massive system is detected is whether we can confidently establish that the system contains an IMBH. As the coalescence of \IMBHB\ systems is likely to be electromagnetically quiet, GW observations will be essential to measuring the parameters of these systems. Our results indicate that advanced GW detectors, using models which include inspiral, merger and ringdown, will be able to constrain the masses of detected \acp{IMBHB}, at least under the assumption of aligned spins. Figure \ref{fig:m1lowerbound} shows the $5\%$ lower bound on the mass of the more massive component $m_1$ as that parameter increases. This indicates that, at a network SNR of 15 or greater, the accuracy of inference will be sufficient to determine at 95\% confidence that a system with non-spinning components does indeed contain an intermediate mass black hole with mass $>100$\,\Msun when the mass of at least one component is $\sim 130$\,\Msun\ or greater.

The SEOBNR reduced order models~\cite{Puerrer:2014fza,Purrer:2015tud} have proven to be instrumental for performing systematic parameter estimation studies with SEOBNR waveforms and provide speedups of up to several orders of magnitude. At the high total masses and low sampling rate used in this study time-domain SEOBNR waveforms are comparatively quick to generate, but the speedup gained from ROM is still very significant. A single waveform evaluation with the ROMs is roughly 700 times faster than the likelihood computed from the time-domain SEOBNRv2 model. Due to overhead costs the overall runtime of the simulations is roughly a factor 50 cheaper than with time-domain SEOBNR waveforms, allowing parameter estimation on sub-day timescales.
\referee{As the waveform models improve further we will be able to analyse a broader range of physical effects, including higher harmonics, with the rapidity the ROMs provide.}

\vspace{0.2in}
\begin{acknowledgements}
JV was supported by STFC grant ST/K005014/1.
MP was supported by STFC grant ST/I001085/1.
IM acknowledges STFC support and the hospitality of the Monash Center for Astrophysics, supported by a Monash Research Acceleration Grant (PI Y.~Levin).
We thank Mark Hannam and Jonathan Gair for useful discussions.
\end{acknowledgements}
\bibliography{cbc-group,IMBH}

\begin{thebibliography}{50}
\expandafter\ifx\csname natexlab\endcsname\relax\def\natexlab#1{#1}\fi
\expandafter\ifx\csname bibnamefont\endcsname\relax
  \def\bibnamefont#1{#1}\fi
\expandafter\ifx\csname bibfnamefont\endcsname\relax
  \def\bibfnamefont#1{#1}\fi
\expandafter\ifx\csname citenamefont\endcsname\relax
  \def\citenamefont#1{#1}\fi
\expandafter\ifx\csname url\endcsname\relax
  \def\url#1{\texttt{#1}}\fi
\expandafter\ifx\csname urlprefix\endcsname\relax\def\urlprefix{URL }\fi
\providecommand{\bibinfo}[2]{#2}
\providecommand{\eprint}[2][]{\url{#2}}

\bibitem[{\citenamefont{{Harry, G.~M.} and {the LIGO Scientific
  Collaboration}}(2010)}]{AdvLIGO}
\bibinfo{author}{\bibnamefont{{Harry, G.~M.}}} \bibnamefont{and}
  \bibinfo{author}{\bibnamefont{{the LIGO Scientific Collaboration}}},
  \bibinfo{journal}{Classical and Quantum Gravity}
  \textbf{\bibinfo{volume}{27}}, \bibinfo{pages}{084006}
  (\bibinfo{year}{2010}).

\bibitem[{\citenamefont{{Virgo Collaboration}}(2009)}]{AdvVirgo}
\bibinfo{author}{\bibnamefont{{Virgo Collaboration}}}, \bibinfo{type}{Virgo
  Technical Report} \bibinfo{number}{VIR-0027A-09} (\bibinfo{year}{2009}),
  \bibinfo{note}{https://tds.ego-gw.it/itf/tds/file.php?callFile=VIR-0027A-09.pdf}.

\bibitem[{\citenamefont{{LIGO Scientific Collaboration}
  et~al.}(2013)\citenamefont{{LIGO Scientific Collaboration}, {Virgo
  Collaboration}, {Aasi}, {Abadie}, {Abbott}, {Abbott}, {Abbott}, {Abernathy},
  {Accadia}, {Acernese} et~al.}}]{scenarios}
\bibinfo{author}{\bibnamefont{{LIGO Scientific Collaboration}}},
  \bibinfo{author}{\bibnamefont{{Virgo Collaboration}}},
  \bibinfo{author}{\bibfnamefont{J.}~\bibnamefont{{Aasi}}},
  \bibinfo{author}{\bibfnamefont{J.}~\bibnamefont{{Abadie}}},
  \bibinfo{author}{\bibfnamefont{B.~P.} \bibnamefont{{Abbott}}},
  \bibinfo{author}{\bibfnamefont{R.}~\bibnamefont{{Abbott}}},
  \bibinfo{author}{\bibfnamefont{T.~D.} \bibnamefont{{Abbott}}},
  \bibinfo{author}{\bibfnamefont{M.}~\bibnamefont{{Abernathy}}},
  \bibinfo{author}{\bibfnamefont{T.}~\bibnamefont{{Accadia}}},
  \bibinfo{author}{\bibfnamefont{F.}~\bibnamefont{{Acernese}}},
  \bibnamefont{et~al.}, \bibinfo{journal}{ArXiv e-prints}
  (\bibinfo{year}{2013}), \eprint{1304.0670}.

\bibitem[{\citenamefont{Abadie et~al.}(2010)}]{ratesdoc}
\bibinfo{author}{\bibfnamefont{J.}~\bibnamefont{Abadie}} \bibnamefont{et~al.}
  (\bibinfo{collaboration}{LIGO Scientific Collaboration and Virgo
  Collaboration}), \bibinfo{journal}{Class. Quantum Grav.}
  \textbf{\bibinfo{volume}{27}}, \bibinfo{pages}{173001}
  (\bibinfo{year}{2010}).

\bibitem[{\citenamefont{{Mandel} and
  {O'Shaughnessy}}(2010)}]{MandelOShaughnessy:2010}
\bibinfo{author}{\bibfnamefont{I.}~\bibnamefont{{Mandel}}} \bibnamefont{and}
  \bibinfo{author}{\bibfnamefont{R.}~\bibnamefont{{O'Shaughnessy}}},
  \bibinfo{journal}{Classical and Quantum Gravity}
  \textbf{\bibinfo{volume}{27}}, \bibinfo{pages}{114007}
  (\bibinfo{year}{2010}), \eprint{0912.1074}.

\bibitem[{\citenamefont{{Miller} and {Colbert}}(2004)}]{MillerColbert:2004}
\bibinfo{author}{\bibfnamefont{M.~C.} \bibnamefont{{Miller}}} \bibnamefont{and}
  \bibinfo{author}{\bibfnamefont{E.~J.~M.} \bibnamefont{{Colbert}}},
  \bibinfo{journal}{International Journal of Modern Physics D}
  \textbf{\bibinfo{volume}{13}}, \bibinfo{pages}{1} (\bibinfo{year}{2004}),
  \eprint{arXiv:astro-ph/0308402}.

\bibitem[{\citenamefont{{Pasham} et~al.}(2014)\citenamefont{{Pasham},
  {Strohmayer}, and {Mushotzky}}}]{Pasham:2014}
\bibinfo{author}{\bibfnamefont{D.~R.} \bibnamefont{{Pasham}}},
  \bibinfo{author}{\bibfnamefont{T.~E.} \bibnamefont{{Strohmayer}}},
  \bibnamefont{and} \bibinfo{author}{\bibfnamefont{R.~F.}
  \bibnamefont{{Mushotzky}}}, \bibinfo{journal}{\nat}
  \textbf{\bibinfo{volume}{513}}, \bibinfo{pages}{74} (\bibinfo{year}{2014}).

\bibitem[{\citenamefont{{Madau} and {Rees}}(2001)}]{MadauRees:2001}
\bibinfo{author}{\bibfnamefont{P.}~\bibnamefont{{Madau}}} \bibnamefont{and}
  \bibinfo{author}{\bibfnamefont{M.~J.} \bibnamefont{{Rees}}},
  \bibinfo{journal}{\apjl} \textbf{\bibinfo{volume}{551}}, \bibinfo{pages}{L27}
  (\bibinfo{year}{2001}).

\bibitem[{\citenamefont{{Portegies Zwart} and {McMillan}}(2002)}]{PZwart:2002}
\bibinfo{author}{\bibfnamefont{S.~F.} \bibnamefont{{Portegies Zwart}}}
  \bibnamefont{and} \bibinfo{author}{\bibfnamefont{S.~L.~W.}
  \bibnamefont{{McMillan}}}, \bibinfo{journal}{\apj}
  \textbf{\bibinfo{volume}{576}}, \bibinfo{pages}{899} (\bibinfo{year}{2002}),
  \eprint{astro-ph/0201055}.

\bibitem[{\citenamefont{{Portegies Zwart} et~al.}(2004)\citenamefont{{Portegies
  Zwart}, {Baumgardt}, {Hut}, {Makino}, and {McMillan}}}]{PZwart:2004}
\bibinfo{author}{\bibfnamefont{S.~F.} \bibnamefont{{Portegies Zwart}}},
  \bibinfo{author}{\bibfnamefont{H.}~\bibnamefont{{Baumgardt}}},
  \bibinfo{author}{\bibfnamefont{P.}~\bibnamefont{{Hut}}},
  \bibinfo{author}{\bibfnamefont{J.}~\bibnamefont{{Makino}}}, \bibnamefont{and}
  \bibinfo{author}{\bibfnamefont{S.~L.~W.} \bibnamefont{{McMillan}}},
  \bibinfo{journal}{\nat} \textbf{\bibinfo{volume}{428}}, \bibinfo{pages}{724}
  (\bibinfo{year}{2004}), \eprint{arXiv:astro-ph/0402622}.

\bibitem[{\citenamefont{{O'Leary} et~al.}(2006)\citenamefont{{O'Leary},
  {Rasio}, {Fregeau}, {Ivanova}, and {O'Shaughnessy}}}]{OLeary:2006}
\bibinfo{author}{\bibfnamefont{R.~M.} \bibnamefont{{O'Leary}}},
  \bibinfo{author}{\bibfnamefont{F.~A.} \bibnamefont{{Rasio}}},
  \bibinfo{author}{\bibfnamefont{J.~M.} \bibnamefont{{Fregeau}}},
  \bibinfo{author}{\bibfnamefont{N.}~\bibnamefont{{Ivanova}}},
  \bibnamefont{and}
  \bibinfo{author}{\bibfnamefont{R.}~\bibnamefont{{O'Shaughnessy}}},
  \bibinfo{journal}{\apj} \textbf{\bibinfo{volume}{637}}, \bibinfo{pages}{937}
  (\bibinfo{year}{2006}), \eprint{astro-ph/0508224}.

\bibitem[{\citenamefont{{Vesperini} et~al.}(2010)\citenamefont{{Vesperini},
  {McMillan}, {D'Ercole}, and {D'Antona}}}]{Vesperini:2010}
\bibinfo{author}{\bibfnamefont{E.}~\bibnamefont{{Vesperini}}},
  \bibinfo{author}{\bibfnamefont{S.~L.~W.} \bibnamefont{{McMillan}}},
  \bibinfo{author}{\bibfnamefont{A.}~\bibnamefont{{D'Ercole}}},
  \bibnamefont{and}
  \bibinfo{author}{\bibfnamefont{F.}~\bibnamefont{{D'Antona}}},
  \bibinfo{journal}{\apjl} \textbf{\bibinfo{volume}{713}}, \bibinfo{pages}{L41}
  (\bibinfo{year}{2010}), \eprint{1003.3470}.

\bibitem[{\citenamefont{{Mandel} et~al.}(2008)\citenamefont{{Mandel}, {Brown},
  {Gair}, and {Miller}}}]{Mandel:2008}
\bibinfo{author}{\bibfnamefont{I.}~\bibnamefont{{Mandel}}},
  \bibinfo{author}{\bibfnamefont{D.~A.} \bibnamefont{{Brown}}},
  \bibinfo{author}{\bibfnamefont{J.~R.} \bibnamefont{{Gair}}},
  \bibnamefont{and} \bibinfo{author}{\bibfnamefont{M.~C.}
  \bibnamefont{{Miller}}}, \bibinfo{journal}{\apj}
  \textbf{\bibinfo{volume}{681}}, \bibinfo{pages}{1431} (\bibinfo{year}{2008}),
  \eprint{0705.0285}.

\bibitem[{\citenamefont{{Fregeau} et~al.}(2006)\citenamefont{{Fregeau},
  {Larson}, {Miller}, {O'Shaughnessy}, and {Rasio}}}]{Fregeau:2006}
\bibinfo{author}{\bibfnamefont{J.~M.} \bibnamefont{{Fregeau}}},
  \bibinfo{author}{\bibfnamefont{S.~L.} \bibnamefont{{Larson}}},
  \bibinfo{author}{\bibfnamefont{M.~C.} \bibnamefont{{Miller}}},
  \bibinfo{author}{\bibfnamefont{R.}~\bibnamefont{{O'Shaughnessy}}},
  \bibnamefont{and} \bibinfo{author}{\bibfnamefont{F.~A.}
  \bibnamefont{{Rasio}}}, \bibinfo{journal}{Astrophysical Journal Letters}
  \textbf{\bibinfo{volume}{646}}, \bibinfo{pages}{L135} (\bibinfo{year}{2006}),
  \eprint{astro-ph/0605732}.

\bibitem[{\citenamefont{{Amaro-Seoane} and {Freitag}}(2006)}]{Amaro:2006imbh}
\bibinfo{author}{\bibfnamefont{P.}~\bibnamefont{{Amaro-Seoane}}}
  \bibnamefont{and}
  \bibinfo{author}{\bibfnamefont{M.}~\bibnamefont{{Freitag}}},
  \bibinfo{journal}{\apjl} \textbf{\bibinfo{volume}{653}}, \bibinfo{pages}{L53}
  (\bibinfo{year}{2006}), \eprint{astro-ph/0610478}.

\bibitem[{\citenamefont{{Amaro-Seoane} and
  {Santamar{\'{\i}}a}}(2010)}]{AmaroSeoaneSantamaria:2009}
\bibinfo{author}{\bibfnamefont{P.}~\bibnamefont{{Amaro-Seoane}}}
  \bibnamefont{and}
  \bibinfo{author}{\bibfnamefont{L.}~\bibnamefont{{Santamar{\'{\i}}a}}},
  \bibinfo{journal}{\apj} \textbf{\bibinfo{volume}{722}}, \bibinfo{pages}{1197}
  (\bibinfo{year}{2010}), \eprint{0910.0254}.

\bibitem[{\citenamefont{{Belczynski} et~al.}(2014)\citenamefont{{Belczynski},
  {Buonanno}, {Cantiello}, {Fryer}, {Holz}, {Mandel}, {Miller}, and
  {Walczak}}}]{Belczynski:2014VMS}
\bibinfo{author}{\bibfnamefont{K.}~\bibnamefont{{Belczynski}}},
  \bibinfo{author}{\bibfnamefont{A.}~\bibnamefont{{Buonanno}}},
  \bibinfo{author}{\bibfnamefont{M.}~\bibnamefont{{Cantiello}}},
  \bibinfo{author}{\bibfnamefont{C.~L.} \bibnamefont{{Fryer}}},
  \bibinfo{author}{\bibfnamefont{D.~E.} \bibnamefont{{Holz}}},
  \bibinfo{author}{\bibfnamefont{I.}~\bibnamefont{{Mandel}}},
  \bibinfo{author}{\bibfnamefont{M.~C.} \bibnamefont{{Miller}}},
  \bibnamefont{and}
  \bibinfo{author}{\bibfnamefont{M.}~\bibnamefont{{Walczak}}},
  \bibinfo{journal}{\apj} \textbf{\bibinfo{volume}{789}}, \bibinfo{eid}{120}
  (\bibinfo{year}{2014}), \eprint{1403.0677}.

\bibitem[{\citenamefont{{Punturo} et~al.}(2010)\citenamefont{{Punturo},
  {Abernathy}, {Acernese}, {Allen}, {Andersson}, {Arun}, {Barone}, {Barr}
  et~al.}}]{ET}
\bibinfo{author}{\bibfnamefont{M.}~\bibnamefont{{Punturo}}},
  \bibinfo{author}{\bibfnamefont{M.}~\bibnamefont{{Abernathy}}},
  \bibinfo{author}{\bibfnamefont{F.}~\bibnamefont{{Acernese}}},
  \bibinfo{author}{\bibfnamefont{B.}~\bibnamefont{{Allen}}},
  \bibinfo{author}{\bibfnamefont{N.}~\bibnamefont{{Andersson}}},
  \bibinfo{author}{\bibfnamefont{K.}~\bibnamefont{{Arun}}},
  \bibinfo{author}{\bibfnamefont{F.}~\bibnamefont{{Barone}}},
  \bibinfo{author}{\bibfnamefont{B.}~\bibnamefont{{Barr}}},
  \bibnamefont{et~al.}, \bibinfo{journal}{Classical and Quantum Gravity}
  \textbf{\bibinfo{volume}{27}}, \bibinfo{eid}{084007} (\bibinfo{year}{2010}).

\bibitem[{\citenamefont{{Gair} et~al.}(2011)\citenamefont{{Gair}, {Mandel},
  {Miller}, and {Volonteri}}}]{Gair:2009ETrev}
\bibinfo{author}{\bibfnamefont{J.~R.} \bibnamefont{{Gair}}},
  \bibinfo{author}{\bibfnamefont{I.}~\bibnamefont{{Mandel}}},
  \bibinfo{author}{\bibfnamefont{M.~C.} \bibnamefont{{Miller}}},
  \bibnamefont{and}
  \bibinfo{author}{\bibfnamefont{M.}~\bibnamefont{{Volonteri}}},
  \bibinfo{journal}{General Relativity and Gravitation}
  \textbf{\bibinfo{volume}{43}}, \bibinfo{pages}{485} (\bibinfo{year}{2011}),
  \eprint{0907.5450}.

\bibitem[{\citenamefont{{Sesana} et~al.}(2009)\citenamefont{{Sesana}, {Gair},
  {Mandel}, and {Vecchio}}}]{Sesana:2009ET}
\bibinfo{author}{\bibfnamefont{A.}~\bibnamefont{{Sesana}}},
  \bibinfo{author}{\bibfnamefont{J.}~\bibnamefont{{Gair}}},
  \bibinfo{author}{\bibfnamefont{I.}~\bibnamefont{{Mandel}}}, \bibnamefont{and}
  \bibinfo{author}{\bibfnamefont{A.}~\bibnamefont{{Vecchio}}},
  \bibinfo{journal}{\apjl} \textbf{\bibinfo{volume}{698}},
  \bibinfo{pages}{L129} (\bibinfo{year}{2009}), \eprint{0903.4177}.

\bibitem[{\citenamefont{{Gair} et~al.}(2009)\citenamefont{{Gair}, {Mandel},
  {Sesana}, and {Vecchio}}}]{Gair:2009ET}
\bibinfo{author}{\bibfnamefont{J.~R.} \bibnamefont{{Gair}}},
  \bibinfo{author}{\bibfnamefont{I.}~\bibnamefont{{Mandel}}},
  \bibinfo{author}{\bibfnamefont{A.}~\bibnamefont{{Sesana}}}, \bibnamefont{and}
  \bibinfo{author}{\bibfnamefont{A.}~\bibnamefont{{Vecchio}}},
  \bibinfo{journal}{Classical and Quantum Gravity}
  \textbf{\bibinfo{volume}{26}}, \bibinfo{pages}{204009}
  (\bibinfo{year}{2009}), \eprint{0907.3292}.

\bibitem[{\citenamefont{{Brown} et~al.}(2007)\citenamefont{{Brown}, {Brink},
  {Fang}, {Gair}, {Li}, {Lovelace}, {Mandel}, and {Thorne}}}]{Brown:2007}
\bibinfo{author}{\bibfnamefont{D.~A.} \bibnamefont{{Brown}}},
  \bibinfo{author}{\bibfnamefont{J.}~\bibnamefont{{Brink}}},
  \bibinfo{author}{\bibfnamefont{H.}~\bibnamefont{{Fang}}},
  \bibinfo{author}{\bibfnamefont{J.~R.} \bibnamefont{{Gair}}},
  \bibinfo{author}{\bibfnamefont{C.}~\bibnamefont{{Li}}},
  \bibinfo{author}{\bibfnamefont{G.}~\bibnamefont{{Lovelace}}},
  \bibinfo{author}{\bibfnamefont{I.}~\bibnamefont{{Mandel}}}, \bibnamefont{and}
  \bibinfo{author}{\bibfnamefont{K.~S.} \bibnamefont{{Thorne}}},
  \bibinfo{journal}{\prl} \textbf{\bibinfo{volume}{99}},
  \bibinfo{pages}{201102} (\bibinfo{year}{2007}), \eprint{arXiv:gr-qc/0612060}.

\bibitem[{\citenamefont{{Rodriguez} et~al.}(2012)\citenamefont{{Rodriguez},
  {Mandel}, and {Gair}}}]{Rodriguez:2012}
\bibinfo{author}{\bibfnamefont{C.~L.} \bibnamefont{{Rodriguez}}},
  \bibinfo{author}{\bibfnamefont{I.}~\bibnamefont{{Mandel}}}, \bibnamefont{and}
  \bibinfo{author}{\bibfnamefont{J.~R.} \bibnamefont{{Gair}}},
  \bibinfo{journal}{\prd} \textbf{\bibinfo{volume}{85}}, \bibinfo{eid}{062002}
  (\bibinfo{year}{2012}), \eprint{1112.1404}.

\bibitem[{\citenamefont{{Abadie} et~al.}(2012)\citenamefont{{Abadie}, {Abbott},
  {Abbott}, {Abbott}, {Abernathy}, {Accadia}, {Acernese}, {Adams}, {Adhikari},
  {Affeldt} et~al.}}]{IMBHB:S5}
\bibinfo{author}{\bibfnamefont{J.}~\bibnamefont{{Abadie}}},
  \bibinfo{author}{\bibfnamefont{B.~P.} \bibnamefont{{Abbott}}},
  \bibinfo{author}{\bibfnamefont{R.}~\bibnamefont{{Abbott}}},
  \bibinfo{author}{\bibfnamefont{T.~D.} \bibnamefont{{Abbott}}},
  \bibinfo{author}{\bibfnamefont{M.}~\bibnamefont{{Abernathy}}},
  \bibinfo{author}{\bibfnamefont{T.}~\bibnamefont{{Accadia}}},
  \bibinfo{author}{\bibfnamefont{F.}~\bibnamefont{{Acernese}}},
  \bibinfo{author}{\bibfnamefont{C.}~\bibnamefont{{Adams}}},
  \bibinfo{author}{\bibfnamefont{R.}~\bibnamefont{{Adhikari}}},
  \bibinfo{author}{\bibfnamefont{C.}~\bibnamefont{{Affeldt}}},
  \bibnamefont{et~al.}, \bibinfo{journal}{\prd} \textbf{\bibinfo{volume}{85}},
  \bibinfo{eid}{102004} (\bibinfo{year}{2012}), \eprint{1201.5999}.

\bibitem[{\citenamefont{{The LIGO Scientific Collaboration}
  et~al.}(2014{\natexlab{a}})\citenamefont{{The LIGO Scientific Collaboration},
  {the Virgo Collaboration}, {Aasi}, {Abbott}, {Abbott}, {Abbott}, {Abernathy},
  {Accadia}, {Acernese}, {Ackley} et~al.}}]{IMBHB:S6}
\bibinfo{author}{\bibnamefont{{The LIGO Scientific Collaboration}}},
  \bibinfo{author}{\bibnamefont{{the Virgo Collaboration}}},
  \bibinfo{author}{\bibfnamefont{J.}~\bibnamefont{{Aasi}}},
  \bibinfo{author}{\bibfnamefont{B.~P.} \bibnamefont{{Abbott}}},
  \bibinfo{author}{\bibfnamefont{R.}~\bibnamefont{{Abbott}}},
  \bibinfo{author}{\bibfnamefont{T.}~\bibnamefont{{Abbott}}},
  \bibinfo{author}{\bibfnamefont{M.~R.} \bibnamefont{{Abernathy}}},
  \bibinfo{author}{\bibfnamefont{T.}~\bibnamefont{{Accadia}}},
  \bibinfo{author}{\bibfnamefont{F.}~\bibnamefont{{Acernese}}},
  \bibinfo{author}{\bibfnamefont{K.}~\bibnamefont{{Ackley}}},
  \bibnamefont{et~al.}, \bibinfo{journal}{ArXiv e-prints}
  (\bibinfo{year}{2014}{\natexlab{a}}), \eprint{1404.2199}.

\bibitem[{\citenamefont{{The LIGO Scientific Collaboration}
  et~al.}(2014{\natexlab{b}})\citenamefont{{The LIGO Scientific Collaboration},
  {the Virgo Collaboration}, {Aasi}, {Abbott}, {Abbott}, {Abbott}, {Abernathy},
  {Acernese}, {Ackley}, {Adams} et~al.}}]{Ringdown:S5S6}
\bibinfo{author}{\bibnamefont{{The LIGO Scientific Collaboration}}},
  \bibinfo{author}{\bibnamefont{{the Virgo Collaboration}}},
  \bibinfo{author}{\bibfnamefont{J.}~\bibnamefont{{Aasi}}},
  \bibinfo{author}{\bibfnamefont{B.~P.} \bibnamefont{{Abbott}}},
  \bibinfo{author}{\bibfnamefont{R.}~\bibnamefont{{Abbott}}},
  \bibinfo{author}{\bibfnamefont{T.}~\bibnamefont{{Abbott}}},
  \bibinfo{author}{\bibfnamefont{M.~R.} \bibnamefont{{Abernathy}}},
  \bibinfo{author}{\bibfnamefont{F.}~\bibnamefont{{Acernese}}},
  \bibinfo{author}{\bibfnamefont{K.}~\bibnamefont{{Ackley}}},
  \bibinfo{author}{\bibfnamefont{C.}~\bibnamefont{{Adams}}},
  \bibnamefont{et~al.}, \bibinfo{journal}{ArXiv e-prints}
  (\bibinfo{year}{2014}{\natexlab{b}}), \eprint{1403.5306}.

\bibitem[{\citenamefont{Buonanno and Damour}(1999)}]{Buonanno:1998gg}
\bibinfo{author}{\bibfnamefont{A.}~\bibnamefont{Buonanno}} \bibnamefont{and}
  \bibinfo{author}{\bibfnamefont{T.}~\bibnamefont{Damour}},
  \bibinfo{journal}{Phys. Rev.} \textbf{\bibinfo{volume}{D59}},
  \bibinfo{pages}{084006} (\bibinfo{year}{1999}), \eprint{gr-qc/9811091}.

\bibitem[{\citenamefont{Buonanno and Damour}(2000)}]{Buonanno:2000ef}
\bibinfo{author}{\bibfnamefont{A.}~\bibnamefont{Buonanno}} \bibnamefont{and}
  \bibinfo{author}{\bibfnamefont{T.}~\bibnamefont{Damour}},
  \bibinfo{journal}{Phys. Rev.} \textbf{\bibinfo{volume}{D62}},
  \bibinfo{pages}{064015} (\bibinfo{year}{2000}).

\bibitem[{\citenamefont{Damour}(2001)}]{Damour:2001tu}
\bibinfo{author}{\bibfnamefont{T.}~\bibnamefont{Damour}},
  \bibinfo{journal}{Phys.Rev.} \textbf{\bibinfo{volume}{D64}},
  \bibinfo{pages}{124013} (\bibinfo{year}{2001}), \eprint{gr-qc/0103018}.

\bibitem[{\citenamefont{Field et~al.}(2014)\citenamefont{Field, Galley,
  Hesthaven, Kaye, and Tiglio}}]{Field:2013cfa}
\bibinfo{author}{\bibfnamefont{S.~E.} \bibnamefont{Field}},
  \bibinfo{author}{\bibfnamefont{C.~R.} \bibnamefont{Galley}},
  \bibinfo{author}{\bibfnamefont{J.~S.} \bibnamefont{Hesthaven}},
  \bibinfo{author}{\bibfnamefont{J.}~\bibnamefont{Kaye}}, \bibnamefont{and}
  \bibinfo{author}{\bibfnamefont{M.}~\bibnamefont{Tiglio}},
  \bibinfo{journal}{Phys.Rev.} \textbf{\bibinfo{volume}{X4}},
  \bibinfo{pages}{031006} (\bibinfo{year}{2014}), \eprint{1308.3565}.

\bibitem[{\citenamefont{P{\"u}rrer}(2014)}]{Puerrer:2014fza}
\bibinfo{author}{\bibfnamefont{M.}~\bibnamefont{P{\"u}rrer}},
  \bibinfo{journal}{Class.Quant.Grav.} \textbf{\bibinfo{volume}{31}},
  \bibinfo{pages}{195010} (\bibinfo{year}{2014}), \eprint{1402.4146}.

\bibitem[{\citenamefont{P{\"u}rrer}(2015)}]{Purrer:2015tud}
\bibinfo{author}{\bibfnamefont{M.}~\bibnamefont{P{\"u}rrer}},
  \bibinfo{journal}{ArXiv e-prints}  (\bibinfo{year}{2015}),
  \eprint{1512.02248}.

\bibitem[{\citenamefont{Taracchini et~al.}(2012)\citenamefont{Taracchini, Pan,
  Buonanno, Barausse, Boyle, Chu, Lovelace, Pfeiffer, and
  Scheel}}]{Taracchini:2012ig}
\bibinfo{author}{\bibfnamefont{A.}~\bibnamefont{Taracchini}},
  \bibinfo{author}{\bibfnamefont{Y.}~\bibnamefont{Pan}},
  \bibinfo{author}{\bibfnamefont{A.}~\bibnamefont{Buonanno}},
  \bibinfo{author}{\bibfnamefont{E.}~\bibnamefont{Barausse}},
  \bibinfo{author}{\bibfnamefont{M.}~\bibnamefont{Boyle}},
  \bibinfo{author}{\bibfnamefont{T.}~\bibnamefont{Chu}},
  \bibinfo{author}{\bibfnamefont{G.}~\bibnamefont{Lovelace}},
  \bibinfo{author}{\bibfnamefont{H.~P.} \bibnamefont{Pfeiffer}},
  \bibnamefont{and} \bibinfo{author}{\bibfnamefont{M.~A.}
  \bibnamefont{Scheel}}, \bibinfo{journal}{Phys. Rev.}
  \textbf{\bibinfo{volume}{D86}}, \bibinfo{pages}{024011}
  (\bibinfo{year}{2012}), \eprint{1202.0790}.

\bibitem[{\citenamefont{Taracchini et~al.}(2014)\citenamefont{Taracchini,
  Buonanno, Pan, Hinderer, Boyle et~al.}}]{Taracchini:2013rva}
\bibinfo{author}{\bibfnamefont{A.}~\bibnamefont{Taracchini}},
  \bibinfo{author}{\bibfnamefont{A.}~\bibnamefont{Buonanno}},
  \bibinfo{author}{\bibfnamefont{Y.}~\bibnamefont{Pan}},
  \bibinfo{author}{\bibfnamefont{T.}~\bibnamefont{Hinderer}},
  \bibinfo{author}{\bibfnamefont{M.}~\bibnamefont{Boyle}},
  \bibnamefont{et~al.}, \bibinfo{journal}{Phys.Rev.}
  \textbf{\bibinfo{volume}{D89}}, \bibinfo{pages}{061502}
  (\bibinfo{year}{2014}), \eprint{1311.2544}.

\bibitem[{LALSuite()}]{LAL-web}
LALSuite, \bibinfo{howpublished}{{The LIGO Scientific Collaboration, Lsc
  algorithm library (lal)} --
  https://www.lsc-group.phys.uwm.edu/daswg/projects/lalsuite.html}.

\bibitem[{\citenamefont{Barsotti and Fritschel}(2014)}]{LIGO-T1200307}
\bibinfo{author}{\bibfnamefont{L.}~\bibnamefont{Barsotti}} \bibnamefont{and}
  \bibinfo{author}{\bibfnamefont{P.}~\bibnamefont{Fritschel}},
  \bibinfo{type}{Tech. Rep.} \bibinfo{number}{{T1200307}},
  \bibinfo{institution}{{The LIGO Scientific Collaboration and the Virgo
  Collaboration}} (\bibinfo{year}{2014}),
  \bibinfo{note}{\url{https://dcc.ligo.org/LIGO-T1200307/public}}.

\bibitem[{\citenamefont{{Manzotti} and {Dietz}}(2012)}]{ManzottiDietz:2012}
\bibinfo{author}{\bibfnamefont{A.}~\bibnamefont{{Manzotti}}} \bibnamefont{and}
  \bibinfo{author}{\bibfnamefont{A.}~\bibnamefont{{Dietz}}},
  \bibinfo{journal}{ArXiv e-prints}  (\bibinfo{year}{2012}),
  \eprint{1202.4031},
  \urlprefix\url{http://adsabs.harvard.edu/abs/2012arXiv1202.4031M}.

\bibitem[{\citenamefont{{Berry} et~al.}(2015)\citenamefont{{Berry}, {Mandel},
  {Middleton}, {Singer}, {Urban}, {Vecchio}, {Vitale}, {Cannon}, {Farr}, {Farr}
  et~al.}}]{Berry:2015}
\bibinfo{author}{\bibfnamefont{C.~P.~L.} \bibnamefont{{Berry}}},
  \bibinfo{author}{\bibfnamefont{I.}~\bibnamefont{{Mandel}}},
  \bibinfo{author}{\bibfnamefont{H.}~\bibnamefont{{Middleton}}},
  \bibinfo{author}{\bibfnamefont{L.~P.} \bibnamefont{{Singer}}},
  \bibinfo{author}{\bibfnamefont{A.~L.} \bibnamefont{{Urban}}},
  \bibinfo{author}{\bibfnamefont{A.}~\bibnamefont{{Vecchio}}},
  \bibinfo{author}{\bibfnamefont{S.}~\bibnamefont{{Vitale}}},
  \bibinfo{author}{\bibfnamefont{K.}~\bibnamefont{{Cannon}}},
  \bibinfo{author}{\bibfnamefont{B.}~\bibnamefont{{Farr}}},
  \bibinfo{author}{\bibfnamefont{W.~M.} \bibnamefont{{Farr}}},
  \bibnamefont{et~al.}, \bibinfo{journal}{\apj} \textbf{\bibinfo{volume}{804}},
  \bibinfo{eid}{114} (\bibinfo{year}{2015}), \eprint{1411.6934}.

\bibitem[{\citenamefont{Veitch et~al.}(2015)\citenamefont{Veitch, Raymond,
  Farr, Farr, Graff, Vitale, Aylott, Blackburn, Christensen, Coughlin
  et~al.}}]{LALInference}
\bibinfo{author}{\bibfnamefont{J.}~\bibnamefont{Veitch}},
  \bibinfo{author}{\bibfnamefont{V.}~\bibnamefont{Raymond}},
  \bibinfo{author}{\bibfnamefont{B.}~\bibnamefont{Farr}},
  \bibinfo{author}{\bibfnamefont{W.}~\bibnamefont{Farr}},
  \bibinfo{author}{\bibfnamefont{P.}~\bibnamefont{Graff}},
  \bibinfo{author}{\bibfnamefont{S.}~\bibnamefont{Vitale}},
  \bibinfo{author}{\bibfnamefont{B.}~\bibnamefont{Aylott}},
  \bibinfo{author}{\bibfnamefont{K.}~\bibnamefont{Blackburn}},
  \bibinfo{author}{\bibfnamefont{N.}~\bibnamefont{Christensen}},
  \bibinfo{author}{\bibfnamefont{M.}~\bibnamefont{Coughlin}},
  \bibnamefont{et~al.}, \bibinfo{journal}{Phys. Rev. D}
  \textbf{\bibinfo{volume}{91}}, \bibinfo{pages}{042003}
  (\bibinfo{year}{2015}).

\bibitem[{\citenamefont{Littenberg et~al.}(2013)\citenamefont{Littenberg,
  Baker, Buonanno, and Kelly}}]{Littenberg:2012uj}
\bibinfo{author}{\bibfnamefont{T.~B.} \bibnamefont{Littenberg}},
  \bibinfo{author}{\bibfnamefont{J.~G.} \bibnamefont{Baker}},
  \bibinfo{author}{\bibfnamefont{A.}~\bibnamefont{Buonanno}}, \bibnamefont{and}
  \bibinfo{author}{\bibfnamefont{B.~J.} \bibnamefont{Kelly}},
  \bibinfo{journal}{Phys.Rev.} \textbf{\bibinfo{volume}{D87}},
  \bibinfo{pages}{104003} (\bibinfo{year}{2013}), \eprint{1210.0893}.

\bibitem[{\citenamefont{Varma et~al.}(2014)\citenamefont{Varma, Ajith, Husa,
  Bustillo, Hannam et~al.}}]{Varma:2014jxa}
\bibinfo{author}{\bibfnamefont{V.}~\bibnamefont{Varma}},
  \bibinfo{author}{\bibfnamefont{P.}~\bibnamefont{Ajith}},
  \bibinfo{author}{\bibfnamefont{S.}~\bibnamefont{Husa}},
  \bibinfo{author}{\bibfnamefont{J.~C.} \bibnamefont{Bustillo}},
  \bibinfo{author}{\bibfnamefont{M.}~\bibnamefont{Hannam}},
  \bibnamefont{et~al.}, \bibinfo{journal}{Phys.Rev.}
  \textbf{\bibinfo{volume}{D90}}, \bibinfo{pages}{124004}
  (\bibinfo{year}{2014}), \eprint{1409.2349}.

\bibitem[{\citenamefont{{Graff} et~al.}(2015)\citenamefont{{Graff}, {Buonanno},
  and {Sathyaprakash}}}]{Graff:2015}
\bibinfo{author}{\bibfnamefont{P.~B.} \bibnamefont{{Graff}}},
  \bibinfo{author}{\bibfnamefont{A.}~\bibnamefont{{Buonanno}}},
  \bibnamefont{and} \bibinfo{author}{\bibfnamefont{B.~S.}
  \bibnamefont{{Sathyaprakash}}}, \bibinfo{journal}{ArXiv e-prints}
  (\bibinfo{year}{2015}), \eprint{1504.04766}.

\bibitem[{\citenamefont{{Sidery} et~al.}(2014)\citenamefont{{Sidery}, {Aylott},
  {Christensen}, {Farr}, {Farr}, {Feroz}, {Gair}, {Grover}, {Graff}, {Hanna}
  et~al.}}]{Sidery:2013}
\bibinfo{author}{\bibfnamefont{T.}~\bibnamefont{{Sidery}}},
  \bibinfo{author}{\bibfnamefont{B.}~\bibnamefont{{Aylott}}},
  \bibinfo{author}{\bibfnamefont{N.}~\bibnamefont{{Christensen}}},
  \bibinfo{author}{\bibfnamefont{B.}~\bibnamefont{{Farr}}},
  \bibinfo{author}{\bibfnamefont{W.}~\bibnamefont{{Farr}}},
  \bibinfo{author}{\bibfnamefont{F.}~\bibnamefont{{Feroz}}},
  \bibinfo{author}{\bibfnamefont{J.}~\bibnamefont{{Gair}}},
  \bibinfo{author}{\bibfnamefont{K.}~\bibnamefont{{Grover}}},
  \bibinfo{author}{\bibfnamefont{P.}~\bibnamefont{{Graff}}},
  \bibinfo{author}{\bibfnamefont{C.}~\bibnamefont{{Hanna}}},
  \bibnamefont{et~al.}, \bibinfo{journal}{\prd} \textbf{\bibinfo{volume}{89}},
  \bibinfo{eid}{084060} (\bibinfo{year}{2014}), \eprint{1312.6013}.

\bibitem[{\citenamefont{{Berti} et~al.}(2006)\citenamefont{{Berti}, {Cardoso},
  and {Will}}}]{Berti:2006}
\bibinfo{author}{\bibfnamefont{E.}~\bibnamefont{{Berti}}},
  \bibinfo{author}{\bibfnamefont{V.}~\bibnamefont{{Cardoso}}},
  \bibnamefont{and} \bibinfo{author}{\bibfnamefont{C.~M.}
  \bibnamefont{{Will}}}, \bibinfo{journal}{\prd} \textbf{\bibinfo{volume}{73}},
  \bibinfo{pages}{064030} (\bibinfo{year}{2006}), \eprint{arXiv:gr-qc/0512160}.

\bibitem[{\citenamefont{{Aasi} et~al.}(2013)\citenamefont{{Aasi}, {Abadie},
  {Abbott}, {Abbott}, {Abbott}, {Abernathy}, {Accadia}, {Acernese}, {Adams},
  {Adams} et~al.}}]{S6PE}
\bibinfo{author}{\bibfnamefont{J.}~\bibnamefont{{Aasi}}},
  \bibinfo{author}{\bibfnamefont{J.}~\bibnamefont{{Abadie}}},
  \bibinfo{author}{\bibfnamefont{B.~P.} \bibnamefont{{Abbott}}},
  \bibinfo{author}{\bibfnamefont{R.}~\bibnamefont{{Abbott}}},
  \bibinfo{author}{\bibfnamefont{T.~D.} \bibnamefont{{Abbott}}},
  \bibinfo{author}{\bibfnamefont{M.}~\bibnamefont{{Abernathy}}},
  \bibinfo{author}{\bibfnamefont{T.}~\bibnamefont{{Accadia}}},
  \bibinfo{author}{\bibfnamefont{F.}~\bibnamefont{{Acernese}}},
  \bibinfo{author}{\bibfnamefont{C.}~\bibnamefont{{Adams}}},
  \bibinfo{author}{\bibfnamefont{T.}~\bibnamefont{{Adams}}},
  \bibnamefont{et~al.}, \bibinfo{journal}{\prd} \textbf{\bibinfo{volume}{88}},
  \bibinfo{eid}{062001} (\bibinfo{year}{2013}), \eprint{1304.1775}.

\bibitem[{\citenamefont{Rodriguez et~al.}(2014)\citenamefont{Rodriguez, Farr,
  Raymond, Farr, Littenberg et~al.}}]{RodriguezEtAl:2013}
\bibinfo{author}{\bibfnamefont{C.~L.} \bibnamefont{Rodriguez}},
  \bibinfo{author}{\bibfnamefont{B.}~\bibnamefont{Farr}},
  \bibinfo{author}{\bibfnamefont{V.}~\bibnamefont{Raymond}},
  \bibinfo{author}{\bibfnamefont{W.~M.} \bibnamefont{Farr}},
  \bibinfo{author}{\bibfnamefont{T.~B.} \bibnamefont{Littenberg}},
  \bibnamefont{et~al.}, \bibinfo{journal}{Astrophys.J.}
  \textbf{\bibinfo{volume}{784}}, \bibinfo{pages}{119} (\bibinfo{year}{2014}),
  \eprint{1309.3273}.

\bibitem[{\citenamefont{{Mandel} et~al.}(2015)\citenamefont{{Mandel}, {Haster},
  {Dominik}, and {Belczynski}}}]{Mandel:2015}
\bibinfo{author}{\bibfnamefont{I.}~\bibnamefont{{Mandel}}},
  \bibinfo{author}{\bibfnamefont{C.-J.} \bibnamefont{{Haster}}},
  \bibinfo{author}{\bibfnamefont{M.}~\bibnamefont{{Dominik}}},
  \bibnamefont{and}
  \bibinfo{author}{\bibfnamefont{K.}~\bibnamefont{{Belczynski}}},
  \bibinfo{journal}{\mnras} \textbf{\bibinfo{volume}{450}},
  \bibinfo{pages}{L85} (\bibinfo{year}{2015}), \eprint{1503.03172}.

\bibitem[{\citenamefont{{Astone} et~al.}(2015)\citenamefont{{Astone},
  {Weinstein}, {Agathos}, {Bejger}, {Christensen}, {Dent}, {Graff}, {Klimenko},
  {Mazzolo}, {Nishizawa} et~al.}}]{2015GReGr..47...11A}
\bibinfo{author}{\bibfnamefont{P.}~\bibnamefont{{Astone}}},
  \bibinfo{author}{\bibfnamefont{A.}~\bibnamefont{{Weinstein}}},
  \bibinfo{author}{\bibfnamefont{M.}~\bibnamefont{{Agathos}}},
  \bibinfo{author}{\bibfnamefont{M.}~\bibnamefont{{Bejger}}},
  \bibinfo{author}{\bibfnamefont{N.}~\bibnamefont{{Christensen}}},
  \bibinfo{author}{\bibfnamefont{T.}~\bibnamefont{{Dent}}},
  \bibinfo{author}{\bibfnamefont{P.}~\bibnamefont{{Graff}}},
  \bibinfo{author}{\bibfnamefont{S.}~\bibnamefont{{Klimenko}}},
  \bibinfo{author}{\bibfnamefont{G.}~\bibnamefont{{Mazzolo}}},
  \bibinfo{author}{\bibfnamefont{A.}~\bibnamefont{{Nishizawa}}},
  \bibnamefont{et~al.}, \bibinfo{journal}{General Relativity and Gravitation}
  \textbf{\bibinfo{volume}{47}}, \bibinfo{pages}{11} (\bibinfo{year}{2015}).

\bibitem[{\citenamefont{Ajith}(2011)}]{Ajith:2011ec}
\bibinfo{author}{\bibfnamefont{P.}~\bibnamefont{Ajith}},
  \bibinfo{journal}{\prd} \textbf{\bibinfo{volume}{84}},
  \bibinfo{pages}{084037} (\bibinfo{year}{2011}), \eprint{1107.1267}.

\bibitem[{\citenamefont{P{\"u}rrer et~al.}(2013)\citenamefont{P{\"u}rrer,
  Hannam, Ajith, and Husa}}]{Purrer:2013ojf}
\bibinfo{author}{\bibfnamefont{M.}~\bibnamefont{P{\"u}rrer}},
  \bibinfo{author}{\bibfnamefont{M.}~\bibnamefont{Hannam}},
  \bibinfo{author}{\bibfnamefont{P.}~\bibnamefont{Ajith}}, \bibnamefont{and}
  \bibinfo{author}{\bibfnamefont{S.}~\bibnamefont{Husa}},
  \bibinfo{journal}{Phys.Rev.} \textbf{\bibinfo{volume}{D88}},
  \bibinfo{pages}{064007} (\bibinfo{year}{2013}), \eprint{1306.2320}.

\end{thebibliography}
\end{document}